\documentclass[12pt, DIV10,a4paper]{article}

\usepackage{color}
\usepackage{float}
\usepackage[bf,small]{caption2}
\usepackage{comment}
\usepackage{amsmath}
\usepackage{bigints}
\usepackage{rotating, booktabs}
\usepackage{amsopn}
\usepackage{amsfonts}
\usepackage{amsthm}
\usepackage{amssymb}
\usepackage{amsbsy}
\usepackage{gensymb}
\usepackage{multirow}
\usepackage{titling}
\usepackage{natbib}
\usepackage{tocbibind}
\usepackage[a4paper,colorlinks,breaklinks,bookmarksopen,bookmarksnumbered]{hyperref}
\usepackage{epsfig,psfrag}
\usepackage{graphicx}
\usepackage{amsmath}
\usepackage{epstopdf}
\usepackage{tabularx}
\usepackage{subfigure}
\usepackage[mathscr]{euscript}
\usepackage[margin=1in]{geometry}
\usepackage{xr-hyper}

\usepackage{amsfonts} 
\usepackage{calc}
\usepackage{titlesec}

\newcommand{\ueq}[1][]{%
  \if\relax\detokenize{#1}\relax
    \sbox0{$\underbrace{=}_{}$}%
    \mathrel{\mathmakebox[\wd0]{=}}
  \else
    \mathrel{\underbrace{=}_{\mathclap{#1}}}
  \fi}

\newcommand{\bzero}{\boldsymbol{0}}

\newcommand {\ctn}{\cite}

\usepackage{url}
\renewcommand{\t}{\ensuremath{\theta}}

\newcommand{\btheta}{\boldsymbol{\theta}}

\newcommand{\bbeta}{\boldsymbol{\beta}}

\newcommand{\bphi}{\boldsymbol{\phi}}

\newcommand{\bxi}{\boldsymbol{\xi}}

\newcommand{\bTheta}{\boldsymbol{\Theta}}
\newcommand{\bgamma}{\boldsymbol{\gamma}}
\newcommand{\bSigma}{\boldsymbol{\Sigma}}

\newcommand{\bmu}{\boldsymbol{\mu}}

\newcommand{\bD}{\boldsymbol{D}}

\newcommand{\bh}{\boldsymbol{h}}
\newcommand{\bH}{\boldsymbol{H}}

\newcommand{\bA}{\boldsymbol{A}}

\newcommand{\bR}{\boldsymbol{R}}

\newcommand{\bs}{\boldsymbol{s}}

\newcommand{\bx}{\boldsymbol{x}}
\newcommand{\bX}{\boldsymbol{X}}
\newcommand{\by}{\boldsymbol{y}}
\newcommand{\bY}{\boldsymbol{Y}}
\newcommand{\bZ}{\boldsymbol{Z}}

\renewcommand{\t}{\ensuremath{\theta}}

\newcommand{\topline}{\hrule height 1pt width \textwidth \vspace*{2pt}}
\newcommand{\botline}{\vspace*{2pt}\hrule height 1pt width \textwidth \vspace*{4pt}}
\newtheorem{algo}{Algorithm} 

\begin{document}

\title{\vspace{-0.8in}
\textbf{IID Sampling from Doubly Intractable Distributions}}
\author{Sourabh Bhattacharya\thanks{
Sourabh Bhattacharya is an Associate Professor in Interdisciplinary Statistical Research Unit, Indian Statistical
Institute, 203, B. T. Road, Kolkata 700108.
Corresponding e-mail: sourabh@isical.ac.in.}}
\date{\vspace{-0.5in}}
\maketitle%
	
\begin{abstract}

Intractable posterior distributions of parameters with intractable normalizing constants depending upon the parameters are known as doubly intractable
posterior distributions. The terminology itself indicates that obtaining Bayesian inference from such posteriors is doubly difficult compared
to traditional intractable posteriors where the normalizing constants are tractable and admit traditional Markov Chain Monte Carlo (MCMC) solutions. 

As can be anticipated, a plethora of MCMC-based methods have originated in the literature to deal with doubly intractable distributions.
Yet, it remains very much unclear if any of the methods can satisfactorily sample from such posteriors, particularly in high-dimensional setups.

In this article, we consider efficient Monte Carlo and importance sampling approximations of the intractable normalizing constant for a few values of the parameters, and Gaussian
process interpolations for the remaining values of the parameters, using the approximations.
We then incorporate this strategy within the exact $iid$ sampling framework developed in \ctn{Bhatta21a} and \ctn{Bhatta21b}, and illustrate
the methodology with simulation experiments comprising a two-dimensional normal-gamma posterior, a two-dimensional Ising model posterior, 
a two-dimensional Strauss process posterior and a $100$-dimensional autologistic model posterior.
In each case we demonstrate great accuracy of our methodology, which is also computationally extremely efficient, often taking only a few minutes
for generating $10,000$ $iid$ realizations on $80$ processors.
\\[2mm]
{\bf Keywords:} {\it Doubly intractable distribution; Importance sampling; Gaussian process; Parallel computing; Perfect sampling; 
Transformation based Markov Chain Monte Carlo.}

\end{abstract}
	
\pagebreak

\section{Introduction}
\label{sec:introduction}

Inference on intractable probability distributions, such as complicated posterior distributions, 
is usually carried on by traditional Markov Chain Monte Carlo (MCMC) sampling methods. 
Indeed, the latter has literally revolutionized Bayesian computation and in principle, has enabled Bayesian inference in various realistic and complex setups
which could not be imagined in the pre-MCMC era. However, issues of practically ascertaining convergence of MCMC algorithms seem to have discouraged the statistical
community from its desirable widespread use, particularly, in high-dimensional setups.

There is an even more challenging setup, where traditional MCMC methods are not even applicable. This is the framework of doubly intractable distributions of parameters,
where the normalizing constant, depending upon the parameters, is itself intractable. Challenging problems provide career opportunities to researchers and in this case
it was no different, vindicated by the emergence of a large number of papers on this topic. Yet, careers of the researchers notwithstanding, from the scientific
perspective it seems to be doubtful if any of the existing methods is reliable enough, particularly, in high dimensions.  

In order to settle the issue of MCMC convergence assessment in arbitrary fixed-dimensional, multimodal and variable-dimensional problems, 
\ctn{Bhatta21a} and \ctn{Bhatta21b} have introduced novel methodologies for simulating perfect, $iid$ realizations from the target distributions.
In this paper, we propose a Gaussian process interpolation method along with efficient Monte Carlo/importance sampling methods for approximating the intractable normalizing
constants of doubly intractable distributions, and embed it in the $iid$ sampling framework of \ctn{Bhatta21a} and \ctn{Bhatta21b}, to obtain a very efficient
$iid$ simulation scheme for doubly intractable distributions. Our simulation experiments with posterior distributions associated with the normal-gamma model,
Ising model, Strauss process model and an autologistic model consisting of $100$ parameters, demonstrate high accuracy and computational efficiency
of our methodology. Indeed, in most cases, our algorithm takes only a few minutes to simulate $10,000$ $iid$ realizations on $80$ processors.

The rest of our article is organized as follows. 
In Section \ref{sec:di_normconst} we provide a briefing on doubly intractable distributions and introduce our idea of approximating the unknown 
normalizing constants by combining Monte Carlo/importance sampling with appropriate Gaussian process interpolation. An overview of the $iid$ sampling idea
of \ctn{Bhatta21a} and \ctn{Bhatta21b} in the doubly intractable context, is provided in Section \ref{sec:idea}.
The complete algorithm for $iid$ sampling from doubly intractable distributions is provided in Section \ref{sec:complete_algo}.
In Section \ref{sec:simstudy} we provide details on four simulation experiments with which we validate our $iid$ sampling algorithm.
Finally, in Section \ref{sec:conclusion}, we summarize our ideas and make concluding remarks.

\section{Doubly intractable distributions and approximating normalizing constants by Gaussian process interpolation}
\label{sec:di_normconst}
Given data $\by=(y_1,\ldots,y_n)$, and prior $\pi(\btheta)$ on the parameter vector $\btheta=(\theta_1,\ldots,\theta_d)^T\in\bTheta$, where
$\bTheta$ is a $d~(\geq 1)$-dimensional parameter space, doubly intractable posteriors admit the following form with respect to some appropriate measure $\nu$
on the data-space $\mathcal X$:
\begin{equation}
	\pi(\btheta|\by)\propto \pi(\btheta)\frac{f(\by|\btheta)}{C(\btheta)},
	\label{eq:di}
\end{equation}
where $f(\by|\btheta)$ is the unnormalized joint density of $\by$ given the parameters, and 
\begin{equation}
	C(\btheta)=\int_{\mathcal X} f(\bx|\btheta)d\nu(\bx) 
\label{eq:normconst}
\end{equation}
is the normalizing constant, assumed to be intractable. Typical cases of (\ref{eq:di}) occur in non-Gaussian Markov random fields, spatial point process models,
random graph models, etc.

Since (\ref{eq:normconst}) depends upon $\btheta$ and is intractable, it is clear that traditional Metropolis-Hastings methods are ineffective, since $C(\btheta)$ does not
cancel in the acceptance ratio. Expectedly, researchers came up with a plethora of ideas to combat this difficulty; see \ctn{Park18} regarding a detailed review and
the relationships among the methods, many of which attempt to bypass the unknown normalizing constant by introducing auxiliary variables, and others attempt
to approximate the normalizing constant by various methods. There is another very popular but perhaps much less reliable class of Approximate Bayesian Computation (ABC) methods
for doubly intractable distributions; see \ctn{Sisson18} for a comprehensive review. 
In our opinion, reliability and accuracy of none of the existing methods is yet convincingly demonstrated, particularly,
in high-dimensional posterior setups.

\subsection{Approximation of the intractable normalizing constant}

In this article, we propose the following method to estimate $C(\btheta)$. We model $Z(\cdot)=\log\left[C(\cdot)\right]$ as a Gaussian process with mean
\begin{equation}
	\mu(\btheta)=\bh(\btheta)^T\bbeta
	\label{eq:mu}
\end{equation}
where $\bh(\btheta)=(1,\btheta^T)^T$ and $\bbeta=(\beta_0,\beta_1,\ldots,\beta_p)^T$,
and covariance of the form 
\begin{equation}
	Cov\left(Z(\btheta_1),Z(\btheta_2)\right)=\sigma^2c(\btheta_1,\btheta_2),
	\label{eq:cov}
\end{equation}
for any $\btheta_1,\btheta_2\in\bTheta$, where $\sigma^2$ is the process variance and $c(\cdot,\cdot)$ is some positive-definite correlation function.
In particular, for our applications, we choose 
\begin{equation}
c(\btheta_1,\btheta_2)=\exp\left\{-(\btheta_1-\btheta_2)^T\bD(\btheta_1-\btheta_2)\right\},
	\label{eq:corr}
\end{equation}
where $\bD$ is a diagonal matrix with positive diagonal elements. In particular, we set $\bD$ to be the identity matrix in all our applications, which turned
out to work very well.

Now, let us assume that $iid$ realizations $\bx^{(k)}$; $k=1,\ldots,N$ are available from some envelope density $g(\bx|\btheta)$ with respect to $\nu$, so that
by the strong law of large numbers,
\begin{equation}
	\hat C(\btheta)=\frac{1}{N}\sum_{k=1}^N\frac{f(\bx^{(k)}|\btheta)}{g(\bx^{(k)}|\btheta)}\longrightarrow	
	C(\btheta)=\int_{\mathcal X}\left[\frac{f(\bx|\btheta)}{g(\bx|\btheta)}\right]g(\bx|\btheta)d\nu(\bx)=E_g\left[\frac{f(\bx|\btheta)}{g(\bx|\btheta)}\right],
	\label{eq:slln}
\end{equation}
almost surely, as $N\rightarrow\infty$, the last expression of (\ref{eq:slln}) being the expectation of $\frac{f(\bx|\btheta)}{g(\bx|\btheta)}$ with respect to
$g(\bx|\btheta)$.

For $i=1,\ldots,K$, let $Z(\tilde\btheta_i)=\log\left[\hat C(\tilde\btheta_i)\right]$, where $\tilde\theta_i$ are $iid$ realizations drawn uniformly from $\bTheta$.
Now given $\bZ=\left(Z(\tilde\btheta_1),\ldots,Z(\tilde\btheta_K)\right)^T$ and $\bH=\left(\bh(\tilde\btheta_1),\ldots,\bh(\tilde\btheta_K)\right)^T$, along with
the prior $\pi\left(\bbeta,\sigma^2\right)\propto\sigma^{-2}$, for any $\btheta\in\bTheta$, the estimate of $Z(\btheta)$ is given by the mean of the posterior
predictive distribution of $Z(\cdot)$ given $\bZ$, having the form
\begin{equation}
	\hat C(\btheta)=E\left[Z(\btheta)|\bZ\right]=\bh(\btheta)^T\hat\bbeta+\bs(\btheta)^T\bR^{-1}(\bZ-\bH\hat\bbeta),
	\label{eq:normconst_gp}
\end{equation}
where 
$\bR$ is the matrix with $(i,j)$-th element $c(\tilde\btheta_i,\tilde\btheta_j)$,  
$\bs(\btheta)=\left(c(\btheta,\tilde\btheta_1),\ldots,c(\btheta,\tilde\btheta_K)\right)^T$, and
\begin{equation}
	\hat\bbeta=\left(\bH\bR^{-1}\bH\right)^{-1}\left(\bH\bR^{-1}\bZ\right).
	\label{eq:beta_hat}
\end{equation}

The most important advantage of (\ref{eq:normconst_gp}) is its computational simplicity. Note that once $\bZ$ is obtained and the diagonal elements of $\bD$
are estimated (fixed), $\hat\bbeta$ given by (\ref{eq:beta_hat}) does not depend upon $\btheta$ and hence needs to be computed only once. Similarly,
$\bR^{-1}(\bZ-\bH\hat\bbeta)$ on the right hand side of (\ref{eq:normconst}) also needs to be computed only once. Indeed, these are far simpler computational 
exercises compared to obtaining importance sampling estimates for every $\btheta$ that features in MCMC or $iid$ sampling from the doubly intractable target distributions, 
which is usually computationally demanding since $N$ must be chosen to be
substantially large. Estimates based solely on importance sampling has been considered in \ctn{Bhatta09} for MCMC inference on posteriors associated with 
doubly intractable circular distributions.

In all our illustrations except the $100$-dimensional autologistic example, we shall choose $K$ to be as small as $100$, and for the autologistic
example, we shall set $K=500$. Both these choices are significantly small and hence computationally extremely fast, 
yet this strategy, in conjunction with the Gaussian process interpolation (\ref{eq:normconst_gp})
ensures that the $iid$ simulations generated from the doubly intractable target distributions are extremely accurate.
Also, computing the importance sampling estimates associated with $\tilde\btheta_1,\ldots,\tilde\btheta_K$ is an embarrassingly parallelizable exercise, which
we fruitfully employ in our applications.

An important remark regarding the effect of this approximation of the unknown normalizing constant, on the desired $iid$ sampling procedure from
doubly intractable distributions, follows. Indeed, given any given parameter value, for sufficiently large $N$ and $K$ (depending upon the parameter value), 
the approximation can be made as accurate as desired, so that realizations generated from the doubly intractable distribution under the exact and approximated
normalizing constants would be the same, provided the same method and uniform random number is used for generating the realizations under the two setups. 
This is of course true for
the $iid$ method of \ctn{Bhatta21a} and \ctn{Bhatta21b} in which we embed our idea of approximating the normalizing constant.
The theoretical arguments in this regard remain the same as those in \ctn{Bhatta21a} for justification of the effect of the approximation of several quantities
using Monte Carlo simulation in the $iid$ sampling strategy.

Unfortunately, a simple google search revealed that essentially the same idea based on Gaussian process has already been proposed by \ctn{Park20}, who also derived some
theoretical results regarding MCMC convergence in this regard. 
Thus, although our Gaussian process idea is not the first of its kind, $iid$ sampling from doubly intractable distributions is of course novel, and makes our current
article original.

In the next section, we provide an overview of the $iid$ sampling idea of \ctn{Bhatta21a}, slightly extended to accommodate 
$iid$ sampling from doubly intractable distributions.

\section{The $iid$ sampling idea}
\label{sec:idea}

For $\btheta=(\theta_1,\ldots,\theta_d)^T\in\mathbb R^d$, the doubly intractable target distribution $\pi(\btheta|\by)$ from which $iid$ realizations are sought, 
can be represented as 
\begin{equation}
	\pi(\btheta|\by)=\sum_{i=1}^{\infty}\pi(\bA_i|\by)\pi_i(\btheta|\by).
	\label{eq:p1}
\end{equation}
In the above, $\bA_i$ are disjoint compact subsets of $\mathbb R^d$ such that $\cup_{i=1}^{\infty}\bA_i=\mathbb R^d$, and
\begin{equation}
	\pi_i(\btheta|\by)=\frac{\pi(\btheta|\by)}{\pi(\bA_i|\by)}I_{\bA_i}(\btheta), 
	\label{eq:p2}
\end{equation}
is the distribution of $\btheta$ given $\by$ restricted on $\bA_i$; $I_{\bA_i}$ being the indicator function of $\bA_i$.
In (\ref{eq:p1}), $\pi(\bA_i|\by)=\int_{\bA_i}\pi(d\btheta|\by)\geq 0$. Obviously, $\sum_{i=1}^{\infty}\pi(\bA_i|\by)=1$.

The idea of generating $iid$ realizations from $\pi(\btheta|\by)$ is to randomly select $\pi_i(\cdot|\by)$ with probability $\pi(\bA_i|\by)$ and then to simulate
from $\pi_i(\cdot|\by)$ by perfect sampling. 

\subsection{Choice of the sets $\bA_i$}
\label{subsec:choice_sets}
For some appropriate $d$-dimensional vector $\bmu$ and $d\times d$ positive definite scale matrix $\bSigma$,
we shall set $\bA_i=\{\btheta:c_{i-1}\leq (\btheta-\bmu)^T\bSigma^{-1}(\btheta-\bmu)\leq c_i\}$ for $i=1,2,\ldots$, where $0=c_0<c_1<c_2<\cdots$. 
Note that $\bA_1=\{\btheta:(\btheta-\bmu)^T\bSigma^{-1}(\btheta-\bmu)\leq c_1\}$, and for $i\geq 2$, 
$\bA_i=\{\btheta:(\btheta-\bmu)^T\bSigma^{-1}(\btheta-\bmu)\leq c_i\}\setminus\cup_{j=1}^{i-1}\bA_j$.
%
The radii of the ellipsoids, $\sqrt{c_i}$; $i\geq 1$, play important role in the efficiency of the underlying perfect simulation procedure, and hence 
must be chosen with care.

The choices of $\bmu$ and $\bSigma$ will be based on MCMC estimates of the mean (if it exists, or co-ordinate-wise median otherwise) 
and covariance structure of $\pi$ (it it exists, or some appropriate scale matrix otherwise).  
With regard to MCMC here, we recommend the transformation based Markov Chain Monte Carlo (TMCMC)
methodology introduced by \ctn{Dutta14} that simultaneously updates all the variables using suitable deterministic transformations of some low-dimensional (often
one-dimensional) random variable defined on some relevant support. The low dimensionality leads to
great improvement in acceptance rates and computational efficiency, while ensuring adequate convergence
properties. Moreover, we also recommend addition of a further deterministic step to TMCMC, in a similar vein as in \ctn{Liu00}, since it often 
enhances the convergence properties of TMCMC.
In this article on sampling from doubly intractable distributions, the acceptance ratio of TMCMC will depend upon the normalizing constant $C(\cdot)$,
which must be estimated by $\hat C(\cdot)$ of the form (\ref{eq:normconst_gp}).

\subsection{Selecting $\pi_i(\cdot|\by)$ and perfect sampling from $\pi_i(\cdot|\by)$}
\label{subsec:ptmcmc}
To simulate perfectly from $\pi(\btheta|\by)$, we first need to select $\pi_i(\cdot|\by)$ with probability proportional to $\pi(\bA_i|\by)$ 
for some $i\geq 1$, and then need to simulate exactly from $\pi_i(\cdot|\by)$.
As shown in \ctn{Bhatta21a}, up to a normalizing constant, $\pi(\bA_i|\by)$ can be approximated arbitrarily accurately by Monte Carlo averaging of samples
drawn uniformly on $\bA_i$. Let $\widehat{\tilde\pi(\bA_i|\by)}$ denote the Monte Carlo estimate, where $\tilde\pi(\bA_i|\by)$ is $\pi(\bA_i|\by)$ 
without the normalizing constant.
As established in \ctn{Bhatta21a}, for perfect sampling it is enough to consider $\widehat{\tilde\pi(\bA_i|\by)}$, instead of the true quantities
$\tilde\pi(\bA_i|\by)$. Note that computation of $\widehat{\tilde\pi(\bA_i|\by)}$ for doubly intractable $\pi(\cdot|\by)$ would require evaluation
of $\hat C(\cdot)$ using (\ref{eq:normconst_gp}).

For $\btheta\in\bA_i$, for any Borel set $\mathbb B$ in the Borel $\sigma$-field of $\mathbb R^d$, 
let $P_i(\btheta,\mathbb B\cap\bA_i)$ denote the corresponding Metropolis-Hastings transition probability for $\pi_i(\cdot|\by)$. 
Also let $Q_i(\mathbb B\cap\bA_i)$ denote the uniform distribution on $\bA_i$ with density 
\begin{equation}
	q_i(\btheta)=\frac{1}{\mathcal L(\bA_i)}I_{\bA_i}(\btheta),
	\label{eq:uniform_density}
\end{equation}
$\mathcal L(\bA_i)$ denoting the Lebesgue measure of $\bA_i$; see \ctn{Bhatta21a} for the exact expression.

Let $\hat s_i$ and $\hat S_i$ denote the minimum and maximum of
$\tilde\pi(\cdot)$ over the Monte Carlo realizations drawn uniformly from $\bA_i$ in the course of approximating $\tilde\pi(\bA_i|\by)$ by $\widehat{\tilde\pi(\bA_i|\by)}$.
Let $\hat p_i=\frac{\hat s_i}{\hat S_i}-\eta_i$, where $\eta_i$ is a sufficiently small positive quantity.
As in our previous works, we shall refer to $\hat p_i$ as the minorization probability for $\pi_i(\cdot|\by)$.
Then for all $\btheta\in\bA_i$, 
$P_i(\btheta,\mathbb B\cap\bA_i)\geq \hat p_i~ Q_i(\mathbb B\cap\bA_i)$ is the minorization inequality 
and
\begin{equation}
	R_i(\btheta,\mathbb B\cap\bA_i)=\frac{P_i(\btheta,\mathbb B\cap\bA_i)-\hat p_i~ Q_i(\mathbb B\cap\bA_i)}{1-\hat p_i}
	\label{eq:split2}
\end{equation}
is the residual distribution.
Perfect simulation from $\pi_i$ can be achieved by the following steps:
\begin{itemize}
\item[(a)] Draw $T_i\sim Geometric(\hat p_i)$ with respect to the mass function
\begin{equation}
P(T_i=t)=\hat p_i (1-\hat p_i)^{t-1};~t=1,2,\ldots.
\label{eq:geo}
\end{equation}
\item[(b)] Draw $\btheta^{(-T_i)}\sim Q_i(\cdot)$.
\item[(c)] Using $\btheta^{(-T_i)}$ as the initial value, carry the chain $\btheta^{(t+1)}\sim R_i(\btheta^{(t)},\cdot)$ 
forward for $t=-T_i,-T_i+1,\ldots,-1$. 
\item[(d)] Report $\btheta^{(0)}$ as a perfect realization from $\pi_i$. 
\end{itemize}
The method of simulating from $R_i(\btheta^{(t)},\cdot)$ is detailed in \ctn{Bhatta21a}.

\subsection{Diffeomorphism}
\label{subsec:diffeo}

It is evident from the perfect sampling step (a) following (\ref{eq:split2}) that small values of $\hat p_i$ would lead to large values of $T_i$, which would
render the underlying perfect sampling algorithm inefficient. To solve this problem, \ctn{Bhatta21a} proposed inversion of a diffeomorphism proposed
in \ctn{Johnson12a} to flatten
the posterior distribution in a way that its infimum and the supremum are reasonably close (so that $\hat p_i$ are adequately large) on all the $\bA_i$.

Briefly, when $\pi(\btheta|\by)$ is of interest, then  
\begin{align}
\pi_{\bgamma}(\bgamma|\by)=\pi\left(h(\bgamma)|\by\right)\left|\mbox{det}~\nabla h(\bgamma)\right|
\label{eq:transformed_target}
\end{align}
is the density of $\bgamma=h^{-1}(\btheta)$,
where $h$ is a diffeomorphism, $\nabla h(\bgamma)$ denotes the gradient of $h$ at $\bgamma$ and $\mbox{det}~\nabla h(\bgamma)$ 
stands for the determinant of the gradient of $h$ at $\bgamma$.

\ctn{Johnson12a} 
define the following isotropic function $h:\mathbb R^d\mapsto\mathbb R^d$:
\begin{equation}
	h(\bgamma)=\left\{\begin{array}{cc}f(\|\bgamma\|)\frac{\bgamma}{\|\bgamma\|}, & \bgamma\neq \bzero\\
		0, & \bgamma=\bzero,
\end{array}\right.
\label{eq:isotropy}
\end{equation}
for some function $f: (0,\infty)\mapsto (0,\infty)$, $\|\cdot\|$ being the Euclidean norm.
\ctn{Johnson12a} consider isotropic diffeomorphisms, that is, functions of the form $h$ where 
both $h$ and $h^{-1}$ are continuously differentiable, with the further property that 
$\mbox{det}~\nabla h$ and  $\mbox{det}~\nabla h^{-1}$ are also continuously 
differentiable. Specifically, if $\pi$ is only sub-exponentially light, then the following form of $f:[0,\infty)\mapsto [0,\infty)$ given by
\begin{equation}
f(x)=\left\{\begin{array}{cc}e^{bx}-\frac{e}{3}, & x>\frac{1}{b}\\
x^3\frac{b^3e}{6}+x\frac{be}{2}, & x\leq \frac{1}{b},
\end{array}\right.
\label{eq:diffeo2}
\end{equation}
where $b>0$, ensures that the transformed density $\pi_{\bgamma}$ of the form (\ref{eq:transformed_target}),
is super-exponentially light.

In contrast with \ctn{Johnson12a}, our purpose is to convert the doubly intractable target distribution $\pi(\cdot|\by)$ 
to some thick-tailed distribution $\pi_{\bgamma}(\cdot|\by)$ whose supremum
and infimum are close. 
This suggests application of the transformation $\bgamma=h(\btheta)$, the inverse of the transformation considered in \ctn{Johnson12a}.  
Consequently, the density of $\bgamma$ becomes
\begin{align}
	\pi_{\bgamma}(\bgamma|\by)=\pi\left(h^{-1}(\bgamma)|\by\right)\left|\mbox{det}~\nabla h(\bgamma)\right|^{-1},
\label{eq:transformed_target2}
\end{align}
where $h$ is the same as (\ref{eq:isotropy}) and $f$ is given by (\ref{eq:diffeo2}).
We apply the same transformation to the uniform proposal density (\ref{eq:uniform_density}), so that the new proposal density becomes
\begin{equation}
	q_i(\bgamma)=\frac{1}{\mathcal L(\bA_i)}I_{\bA_i}(h^{-1}(\bgamma))\left|\mbox{det}~\nabla h(\bgamma)\right|^{-1}.
	\label{eq:proposal2}
\end{equation}
For any set $\bA$, let $h(\bA)=\left\{h(\btheta):\btheta\in\bA\right\}$.
Also, let $s_i=\underset{\bgamma\in h(\bA_i)}{\inf}~\frac{\tilde\pi_{\bgamma}(\bgamma|\by)}{q_i(\bgamma)}$ and 
$S_i=\underset{\bgamma\in h(\bA_i)}{\sup}~\frac{\tilde\pi_{\bgamma}(\bgamma|\by)}{q_i(\bgamma)}$, where $\tilde\pi_{\bgamma}(\bgamma|\by)$ 
is the same as (\ref{eq:transformed_target2})
but without the normalizing constant.
Then, with (\ref{eq:proposal2}) as the proposal density, we have 
\begin{align}
	P_i(\bgamma,h(\mathbb B\cap\bA_i))&\geq\int_{h(\mathbb B\cap\bA_i)}
	\min\left\{1,\frac{\tilde\pi_{\bgamma}(\bgamma'|\by)/q_i(\bgamma')}{\tilde\pi_{\bgamma}(\bgamma|\by)/q_i(\bgamma)}\right\}q_i(\bgamma')d\bgamma'\notag\\
	&\geq p_i~Q_i(h(\mathbb B \cap\bA_i)),\notag
\end{align}
where $p_i=s_i/S_i$ and $Q_i$ is the probability measure corresponding to (\ref{eq:proposal2}).
With $\hat p_i=\hat s_i/\hat S_i-\eta_i$, where $\hat s_i$ and $\hat S_i$ are Monte Carlo estimates of $s_i$ and $S_i$ and $\eta_i>0$ is adequately small, 
the rest of the details remain the same as before with necessary modifications pertaining to the new proposal density (\ref{eq:proposal2}) and the new 
Metropolis-Hastings acceptance ratio
with respect to (\ref{eq:proposal2}) incorporated in the subsequent steps. Once $\bgamma$ is generated from (\ref{eq:transformed_target2}) we
transform it back to $\btheta$ using $\btheta=h^{-1}(\bgamma)$.

\section{The complete algorithm for $iid$ sample generation from doubly intractable $\pi(\cdot|\by)$}
\label{sec:complete_algo}
The algorithm for generating $iid$ realizations from the doubly intractable target distribution $\pi(\cdot|\by)$ 
is the same as Algorithm 1 of \ctn{Bhatta21a}; only evaluation of $\hat C(\cdot)$ using (\ref{eq:normconst_gp}) is necessary for implementing
TMCMC (in order to obtain estimates of $\bmu$ and $\bSigma$), for obtaining $\widehat{\tilde\pi(\bA_i)}$, and for computing 
the acceptance ratios in the perfect sampling algorithm for simulating from $\pi_i(\cdot|\by)$.
Nevertheless, for ready reference, we present our algorithm for $iid$ sampling from doubly intractable distributions as Algorithm \ref{algo:perfect}.

\begin{algo}\label{algo:perfect}\topline
	IID sampling from the doubly intractable target distribution $\pi(\cdot|\by)$ \botline \normalfont \ttfamily
\begin{itemize}
	\item[(1)] Using TMCMC based estimates, obtain $\bmu$ and $\bSigma$ required for the sets $\bA_i$; $i\geq 1$. In the TMCMC acceptance
		ratios, estimate $C(\cdot)$ by (\ref{eq:normconst_gp}). In this regard, compute $\bZ$ as an embarrassingly parallel processing exercise.
	\item[(2)] Fix $M$ to be sufficiently large. 
	\item[(3)] Choose the radii $\sqrt{c_i}$; $i=1,\ldots,M$ appropriately. A general strategy for these selections will be discussed in the context
		of the applications.
	\item[(4)] Compute the Monte Carlo estimates $\widehat{\tilde\pi(\bA_i|\by)}$; $i=1,\ldots,M$, in parallel processors. As required in these computations, 
		estimate $C(\cdot)$ by (\ref{eq:normconst_gp}).
                Thus, $M$ parallel processors will obtain all the estimates simultaneously.
	\item[(5)] Instruct each processor to send its respective estimate to all the other processors.
	\item[(6)] Let $\tilde K$ be the required $iid$ sample size from the target distribution $\pi$. Split the job of obtaining $\tilde K$ $iid$ realizations into 
		parallel processors, each processor scheduled to simulate a single realization at a time. That is, with $\tilde K$ parallel processors, $\tilde K$ realizations
		will be simulated\\ simultaneously. In each processor, do the following:
		\begin{enumerate}	
			\item[(i)] Select $\pi_i(\cdot|\by)$ with probability proportional to $\widehat{\tilde\pi(\bA_i|\by)}$. 
			\item[(ii)]If $i=M$ for any processor, then return to Step (2), increase $M$ to $2M$, and repeat the subsequent steps
				(in Step (4) only $\widehat{\tilde\pi(\bA_i|\by)}$; $i=M+1,\ldots,2M$, need to be computed).		
				Else
				\begin{enumerate}
	                \item[(a)] Draw $T_i\sim Geometric(\hat p_i)$ with respect to (\ref{eq:geo}).
                        \item[(b)] Draw $\btheta^{(-T_i)}\sim Q_i(\cdot)$.
                        \item[(c)] Using $\btheta^{(-T_i)}$ as the initial value, carry the chain $\btheta^{(t+1)}\sim \tilde R_i(\btheta^{(t)},\cdot)$ 
				forward for $t=-T_i,-T_i+1,\ldots,-1$. As required in the acceptance ratios, estimate $C(\cdot)$ by (\ref{eq:normconst_gp}).
			\item[(d)] From the current processor, send $\btheta^{(0)}$ to processor $0$ as a perfect realization from $\pi(\cdot|\by)$.
				\end{enumerate}
		\end{enumerate}
	\item[(7)] Processor $0$ stores the $\tilde K$ $iid$ realizations $\left\{\btheta^{(0)}_1.\ldots,\btheta^{(0)}_{\tilde K}\right\}$ thus generated from the 
		doubly intractable target distribution $\pi(\cdot|\by)$.
\end{itemize}
\rmfamily
\botline
\end{algo}
Diffeomorphism must be employed in the perfect sampling strategy of Step (6) of Algorithm \ref{algo:perfect} if $\hat p_i$ are not adequately large.

\section{Illustrations with simulation experiments}
\label{sec:simstudy}

\subsection{Normal-gamma posterior}
\label{subsec:normal_gamma}

\subsubsection{Data generation}
We simulate $\by=(y_1,\ldots,y_n)$ with $n=10$ from $N(0,1)$, the normal distribution with mean $0$ and variance $1$, and model the data as
$N(\psi,\sigma^2)$. Letting $\tau=\sigma^{-2}$, we consider the following prior for $(\psi,\tau)$: $[\psi|\tau]\sim N\left(\psi_0,\tau\right)$
and $[\tau]\sim\mathcal G(\alpha_0,\beta_0)$, the gamma distribution with mean $\alpha_0/\beta_0$ and variance $\alpha_0/\beta^2_0$, where $\alpha_0>0$ and $\beta_0>0$.
We set $\psi_0=0$, $\alpha_0=\beta_0=1$.

\subsubsection{Unknown normalizing constant and reparameterization}
The normalizing constant associated with the likelihood is $\left(\frac{\tau}{2\pi}\right)^{n/2}$. For illustration, we assume that $\sqrt{\frac{\tau}{2\pi}}$ is unknown. 
Indeed, since we know that the data are $iid$, it is enough to assume that $\sqrt{\frac{\tau}{2\pi}}$, rather than $\left(\frac{\tau}{2\pi}\right)^{n/2}$, is unknown.
For convenience, we reparameterize the non-negative parameter $\tau$ as $\exp(2\phi)$, where $-\infty<\phi<\infty$. Hence, the new parameter set of interest is
$\btheta=(\psi,\phi)$.

\subsubsection{Estimation of the unknown normalizing constants}
\label{subsubsec:normconst_normal_gamma}
To estimate $\sqrt{\frac{\tau}{2\pi}}$, given $\btheta=(\psi,\phi)$, we first consider another sequence of ellipsoids and annuli on the real line $\mathbb R$ of the form
$B_i=\left\{x:r_{X,i-1}\leq \sqrt{\tau}|x-\psi|\leq r_{X,i}\right\}$, for $i\geq 1$, where 
$0=r_{X,0}<r_{X,1}<r_{X,2}<\cdots$ are the radii. Clearly, $B_i$ are disjoint and $\cup_{i=1}^{\infty}=\mathbb R$.
Note that the Lebesgue measure of $B_i$ is $\mathcal L(B_i)=2(r_{X,i}-r_{X,i-1})$.
Then, on each $B_i$, letting $f(\cdot|\btheta)$ denote the unnormalized univariate density of the data-variable $y$, the Monte Carlo estimate of the
unknown normalizing constant is given by
\begin{equation}
	C_i = \mathcal L(B_i)\int_{B_i}f(x|\btheta)\times\frac{1}{\mathcal L(B_i)}dx\approx \mathcal L(B_i)\times\frac{1}{N_i}\sum_{k=1}^{N_i}f(x^{(k)}|\btheta)=\hat C_i,
	\label{eq:C_i}
\end{equation}
where, for sufficiently large $N_i$, $x^{(k)}$; $k=1,\ldots,N_i$ are $iid$ realizations drawn uniformly from $B_i$. Then
\begin{equation}
	\hat C=\sum_{i=1}^{\infty}\hat C_i
	\label{eq:C_hat}
\end{equation}
is the estimate of the normalizing constant of the one-dimensional data density. Since this aggregates the individual Monte Carlo estimates in the narrow partitions $B_i$,
this is far more efficient compared to the traditional Monte Carlo method.

To construct $\bZ$ in this example, we first simulate $\tilde\btheta_k$; $k=1,\ldots,K~(=100)$ independently and uniformly from the ball $\{\btheta:\|\btheta\|\leq 1\}$.
Given these realizations, we then set 
$r_{X,1}=0.1$ and for $i=2,\ldots,M_X~(=100)$, $r_{X,i}=r_{X,1}+0.3\times (i-1)$. Also, we set $N_i=1000$ for each $i$, and obtain the estimates $\hat C$ for each
$\tilde\btheta_k$; $k=1,\ldots,K$, following (\ref{eq:C_i}) and (\ref{eq:C_hat}). Once $\bZ$ is obtained this way, we use (\ref{eq:normconst_gp}) 
to estimate $Z(\btheta)$ for values of $\btheta\notin\left\{\tilde\btheta_1,\ldots,\tilde\btheta_K\right\}$.

\subsubsection{TMCMC based estimation of $\bmu$ and $\bSigma$}
\label{subsubsec:tmcmc_normal_gamma}
We estimated $\mu$ and $\Sigma$ corresponding to $\btheta$ using 
additive TMCMC with the scaling constants set to $0.05$ for both the components of $\btheta$, along with
incorporation of the method for estimating the unknown normalizing constant detailed in Section \ref{subsubsec:normconst_normal_gamma}. 
The further deterministic additive transformation step in the same vein as
in \ctn{Liu00} is also incorporated in our TMCMC implementation. We discarded the first $10^5$ iterations as burn-in, and stored one in every $10$ iterations of the next
$5\times 10^5$ iterations to yield $50,000$ TMCMC realizations from the (pretended) doubly intractable posterior. We further selected one in every $5$ stored realizations
to obtain $10,000$ realizations, which we used to construct the estimates of $\bmu$ and $\bSigma$. The time taken for the entire exercise has been less than a minute
in our C code implementation on $80$ cores of our VMWare provided by Indian Statistical Institute. 
The machine has $2$ TB memory and each core has about $2.8$ GHz CPU speed.
Indeed, all our codes for TMCMC and $iid$ method implementation are written in C using the Message Passing Interface (MPI) protocol for parallel processing. 
For TMCMC, parallelization is carried out only for obtaining $\bZ$ by Monte Carlo.
It is important to remark that our TMCMC algorithm ensured excellent mixing properties.

\subsubsection{Implementation of Algorithm \ref{algo:perfect} for $iid$ sample generation}
For $iid$ sampling, along with the above details, 
we set $\sqrt{c_1}=2.3$ and for $i=2,\ldots,M$, set $\sqrt{c_i}=\sqrt{c_1}+0.02\times (i-1)$. With these choices of the radii, it turned out that $M=100$
was sufficient for this problem. 
With $\tilde N_i=5000$ being the Monte Carlo sample size drawn uniformly from $\bA_i$ to construct the estimators
$\widehat{\tilde\pi(\bA_i|\by)}$ for $i=1,\ldots,M~(=100)$, we implemented Algorithm \ref{algo:perfect} on $80$ cores of our VMWare, 
which took $1$ hour and $34$ minutes for this problem to generate $10,000$ $iid$ realizations from the pretended doubly intractable posterior of $\btheta$.

\subsubsection{Results of our $iid$ sampling method}
\begin{figure}
	\centering
	\subfigure [True and $iid$-based density for $\psi$.]{ \label{fig:psi}
	\includegraphics[width=7.5cm,height=7.5cm]{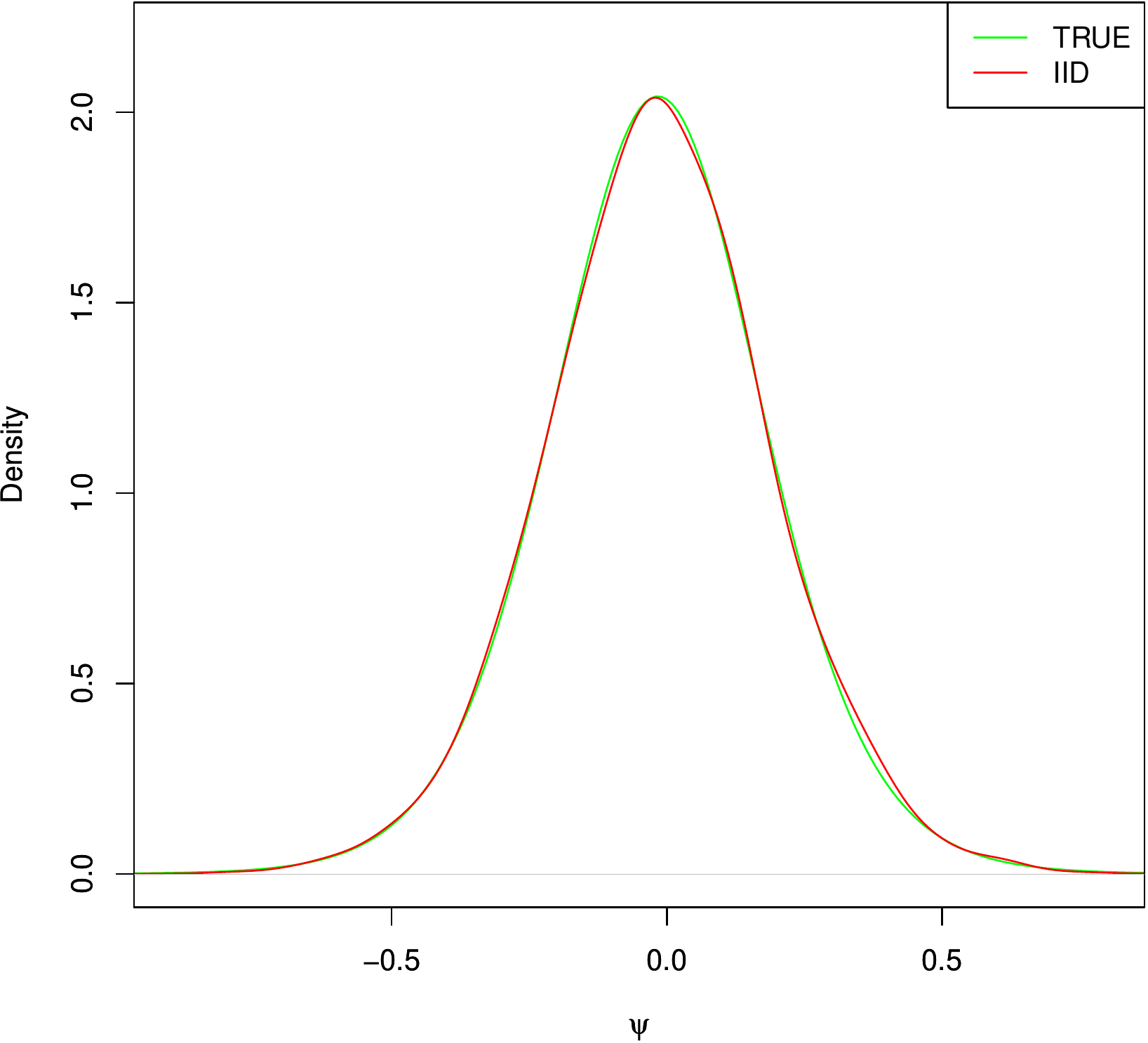}}
	\hspace{2mm}
	\subfigure [True and $iid$-based density for $\tau$.]{ \label{fig:tau}
	\includegraphics[width=7.5cm,height=7.5cm]{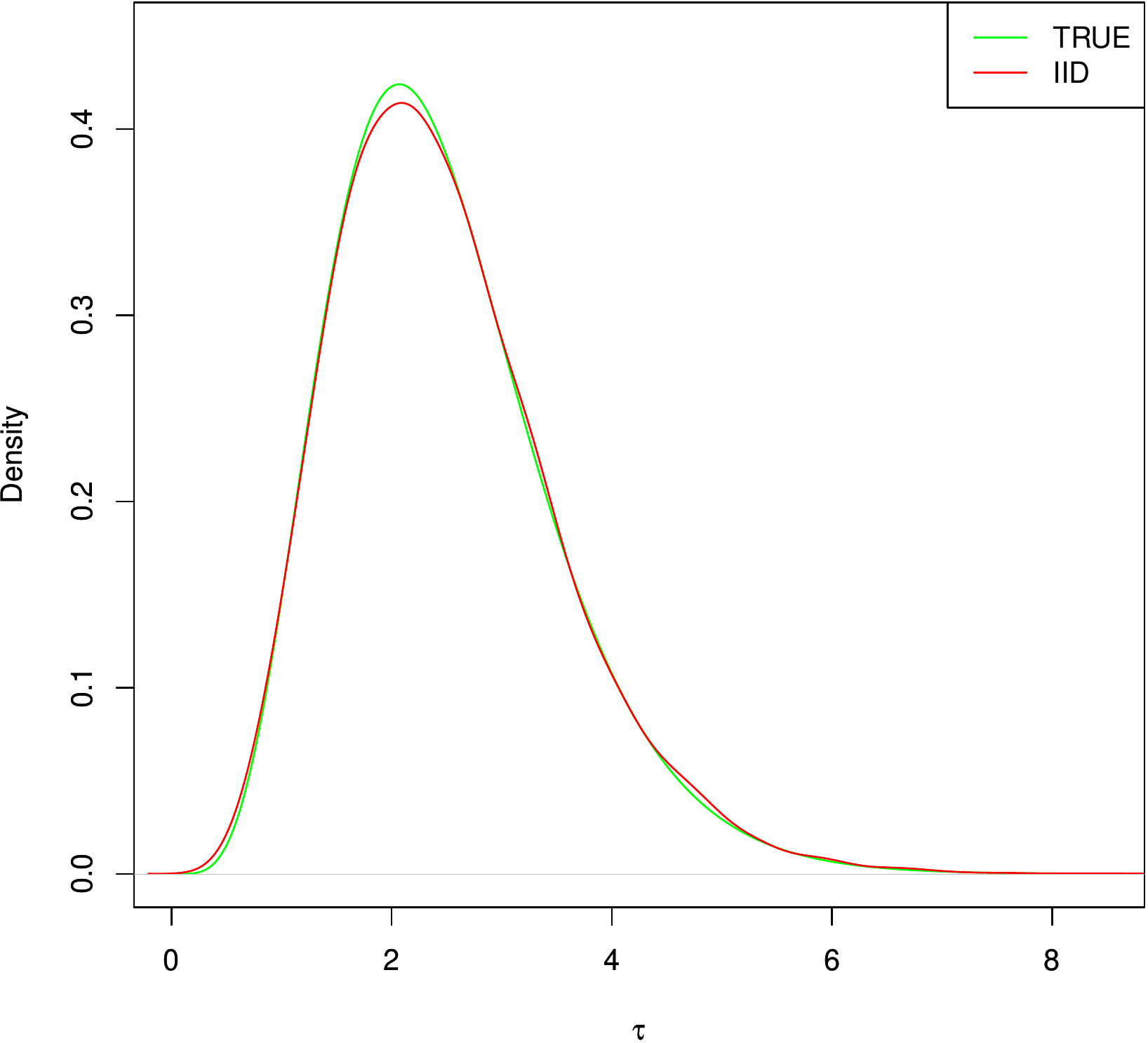}}
	\caption{Normal-gamma simulation experiment. The green and red colours denote the true posterior density and the 
	$iid$ sample based density, respectively.}
	\label{fig:normal_gamma}
\end{figure}

To compare our $iid$ results with the true posterior distributions, let us first provide the details on the true marginal posterior distributions.
In this regard, first let 
\begin{align}
	&\psi_n=\frac{n\bar y+\psi_0}{n+1};\notag\\
	&\alpha_n=2\alpha_0+n;\notag\\
	&\lambda_n=(n+1)\left(\alpha_n+\frac{n}{2}\right)\beta^{-1}_n;\notag\\
	&\beta_n=\beta_0+\frac{ns^2}{2}+\frac{n(\psi_0-\bar y)^2}{2(n+1)},\notag
\end{align}
where $\bar y=n^{-1}\sum_{i=1}^ny_i$ and $s^2=n^{-1}\sum_{i=1}^n(y_i-\bar y)^2$.
The marginal posterior density of $\psi$ is a Student's $t$ distribution of the form
\begin{equation}
	\pi(\psi|\by)=\frac{\Gamma\left(\frac{\alpha_n+1}{2}\right)}{\Gamma\left(\frac{\alpha_n}{2}\right)}\left(\frac{\lambda_n}{\alpha_n\pi}\right)^{1/2}
	\times\left[1+\alpha^{-1}_n\lambda_n(\psi-\psi_n)^2\right]^{-\left(\frac{\alpha_n+1}{2}\right)},
	\label{eq:psi_t}
\end{equation}
and, the marginal posterior distribution of $\tau$ is given by
\begin{align}
	&[\tau|\by]\sim\mathcal G\left(\alpha_0+\frac{n}{2},\beta_n\right).\label{eq:tau_gamma}
\end{align}
Figure \ref{fig:normal_gamma} displays the true marginal posterior distributions of $\psi$ and $\tau$ given by (\ref{eq:psi_t}) and (\ref{eq:tau_gamma}), 
and the corresponding density estimates
based on the $10,000$ $iid$ realizations from the pretended doubly intractable posterior. Close agreement between the true and $iid$-based posteriors vindicate
the accuracy and efficacy of our novel method.

\subsection{Ising model posterior}
\label{subsec:ising}

The well-known Ising model (see, for example, \ctn{Moller06}) models interactions on a binary $m_1\times m_2$ lattice, whose unnormalized density,
with $\by=\left\{y_{i,j}:i=1,\ldots,m_1,j=1,\ldots,m_2\right\}$, is of the form
\begin{equation}
	f(\by|\btheta)=\exp\left(\theta_0V_0+\theta_1V_1\right),
	\label{eq:ising}
\end{equation}
where $\theta_0$, $\theta_1$ are real parameters, and $V_0=\sum_{i=1}^{m_1}\sum_{j=1}^{m_2}y_{ij}$, 
$V_1=\sum_{i=1}^{m_1-1}\sum_{j=1}^{m_2}y_{i,j}y_{i+1,j}+\sum_{i=1}^{m_1}\sum_{j=1}^{m_2-1}y_{i,j}y_{i,j+1}$ are sufficient statistics.
Here $y_{i,j}\in\{-1,1\}$ is the response at location $(i,j)$ of the lattice.

The normalizing constant of (\ref{eq:ising}) depends upon $\theta_0$ and $\theta_1$, but is infeasible to compute, as it requires summation over
$2^{m_1m_2}$ configurations of $\by$.

We assume that {\it a priori}, $\theta_0\sim U(-1,1)$, the uniform distribution of $[-1,1]$ and $\theta_1\sim U(0,1)$.

\subsubsection{Data generation}

With $\theta_0=0.05$ and $\theta_1=0.38$ as the true values, we use Gibbs sampling to generate data from the Ising model with $m_1=m_2=10$. 
For any $(i,j)$, the full conditional distribution
of $y_{i,j}$ given $\by_{-(i,j)}$, the $\by$ vector without $y_{i,j}$, required for Gibbs sampling, is of the following form:
\begin{equation}
	\left[\frac{y_{i,j}+1}{2}\bigg |\by_{-(i,j)}\right]\sim\mathcal B\left(p_{i,j}\right),
	\label{eq:ising_fullcond}
\end{equation}
the Bernoulli distribution with success probability 
\begin{equation*}
p_{i,j}=\frac{\exp\left\{2(\theta_0+\theta_1a_{i,j})\right\}}{1+\exp\left\{2(\theta_0+\theta_1a_{i,j})\right\}},
\end{equation*}
where
\begin{equation*}
	a_{i,j}=y_{i+1,j}I_{\left\{(i+1)\leq m_1\right\}}(i)+y_{i,j+1}I_{\left\{(j+1)\leq m_2\right\}}(j)
	+y_{i-1,j}I_{\left\{(i-1)\geq 1\right\}}(i)+y_{i,j-1}I_{\left\{(j-1)\geq 1\right\}}(j).
\end{equation*}
For Gibbs sampling, we begin with initial values such that $y_{i,j}\in\{-1,1\}$ with probability $1/2$, for all $(i,j)$, and run the Gibbs sampler 
with the full conditionals (\ref{eq:ising_fullcond}) for $10^6$ iterations, and store the realizations of the last iteration as the data $\by$.

\subsubsection{Importance sampling for obtaining $\bZ$}
For importance sampling, we consider the product of the full conditional distributions (\ref{eq:ising_fullcond}) as the envelope distribution, that is,
\begin{equation}
	g(\bx|\by)=\prod_{i=1}^{m_1}\prod_{j=1}^{m_2}\left[\frac{x_{i,j}+1}{2}\bigg |\by_{-(i,j)}\right],
	\label{eq:ising_is}
\end{equation}
from which we generate $10,000$ realizations to estimate the unknown normalizing constant given each value of $(\theta_0,\theta_1)$.

For convenience, as before we reparameterize $\theta_0$ and $\theta_1$ to 
$\theta_0=-1+2\exp\left(\phi_1\right)/\left(1+\exp\left(\phi_1\right)\right)$ and 
$\theta_1=\exp\left(\phi_2\right)/\left(1+\exp\left(\phi_2\right)\right)$, respectively, where $-\infty<\phi_1,\phi_2<\infty$.
We set $\btheta=(\phi_1,\phi_2)$.
As before, we randomly select $\tilde\btheta_1,\ldots,\tilde\btheta_K$ uniformly from the ball of unit radius, with $K=100$. 
For the rest of the values of $\btheta$ we use the Gaussian process interpolation (\ref{eq:normconst_gp}).

\subsubsection{TMCMC implementation to estimate $\bmu$ and $\bSigma$}
With $\btheta=(\phi_1,\phi_2)$, we implement TMCMC to estimate $\bmu$ and $\bSigma$ corresponding to $\btheta$.
The TMCMC details remain the same as in the case of the normal-gamma posterior
as detailed in Section \ref{subsubsec:tmcmc_normal_gamma}, and again exhibited excellent mixing properties.

\subsubsection{Implementation of the $iid$ method}
Here we set $\sqrt{c_1}=2.5$ and for $i=2,\ldots,M~(=100)$, $\sqrt{c_i}=\sqrt{c_1}+0.02\times (i-1)$. 
Again we considered $\tilde N_i=5000$ to be the Monte Carlo sample size drawn uniformly from $\bA_i$ to construct the estimators
$\widehat{\tilde\pi(\bA_i|\by)}$. 

With the above details, implementation of Algorithm \ref{algo:perfect} on $80$ cores of our VMWare took just $2$ minutes for this problem
to generate $10,000$ $iid$ realizations from the doubly intractable posterior of $\btheta$.

\subsubsection{Results}

\begin{figure}
	\centering
	\subfigure [TMCMC and $iid$-based posterior of $\theta_1$.]{ \label{fig:theta0}
	\includegraphics[width=7.5cm,height=7.5cm]{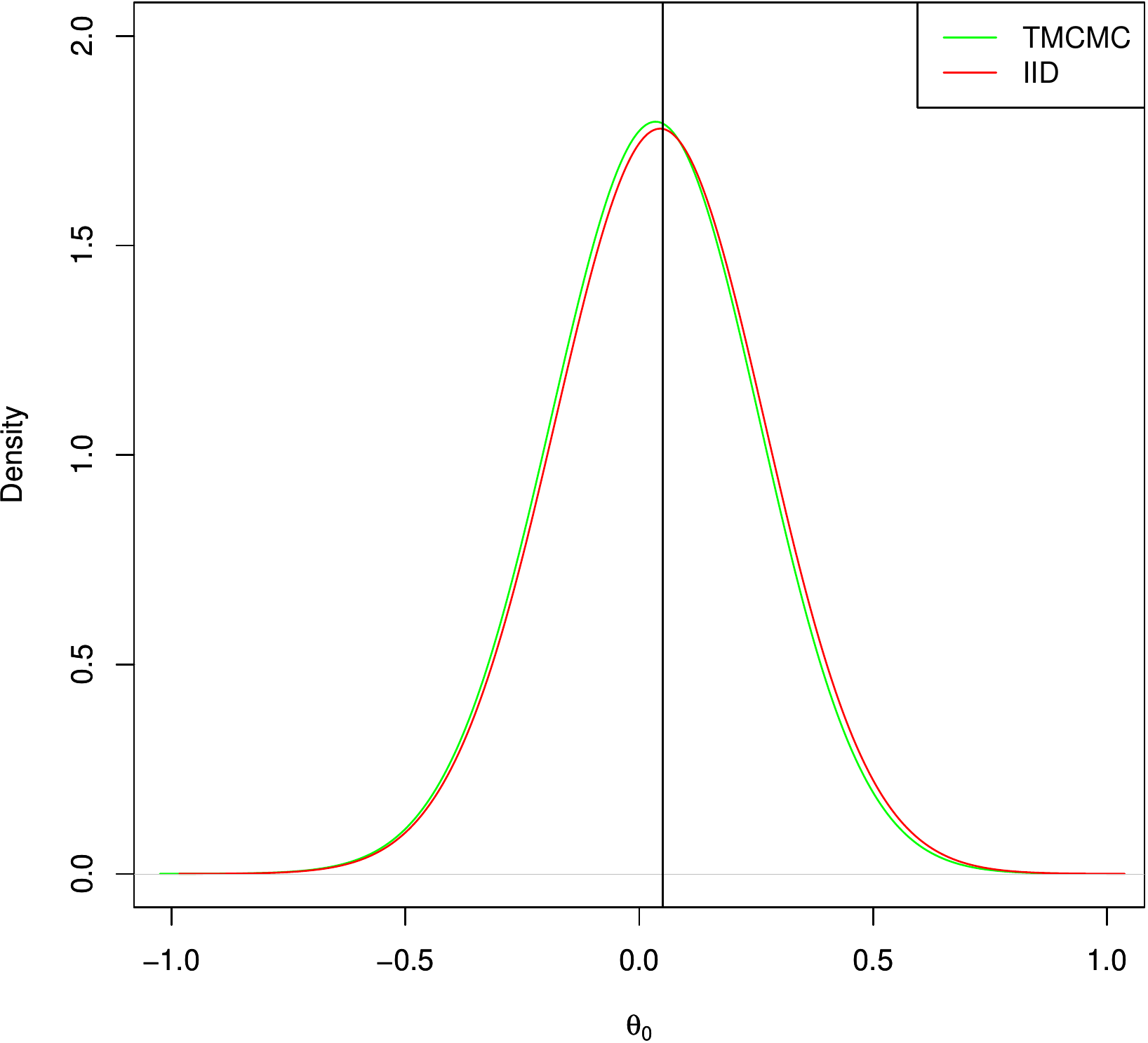}}
	\hspace{2mm}
	\subfigure [TMCMC and $iid$-based posterior of $\theta_2$.]{ \label{fig:theta1}
	\includegraphics[width=7.5cm,height=7.5cm]{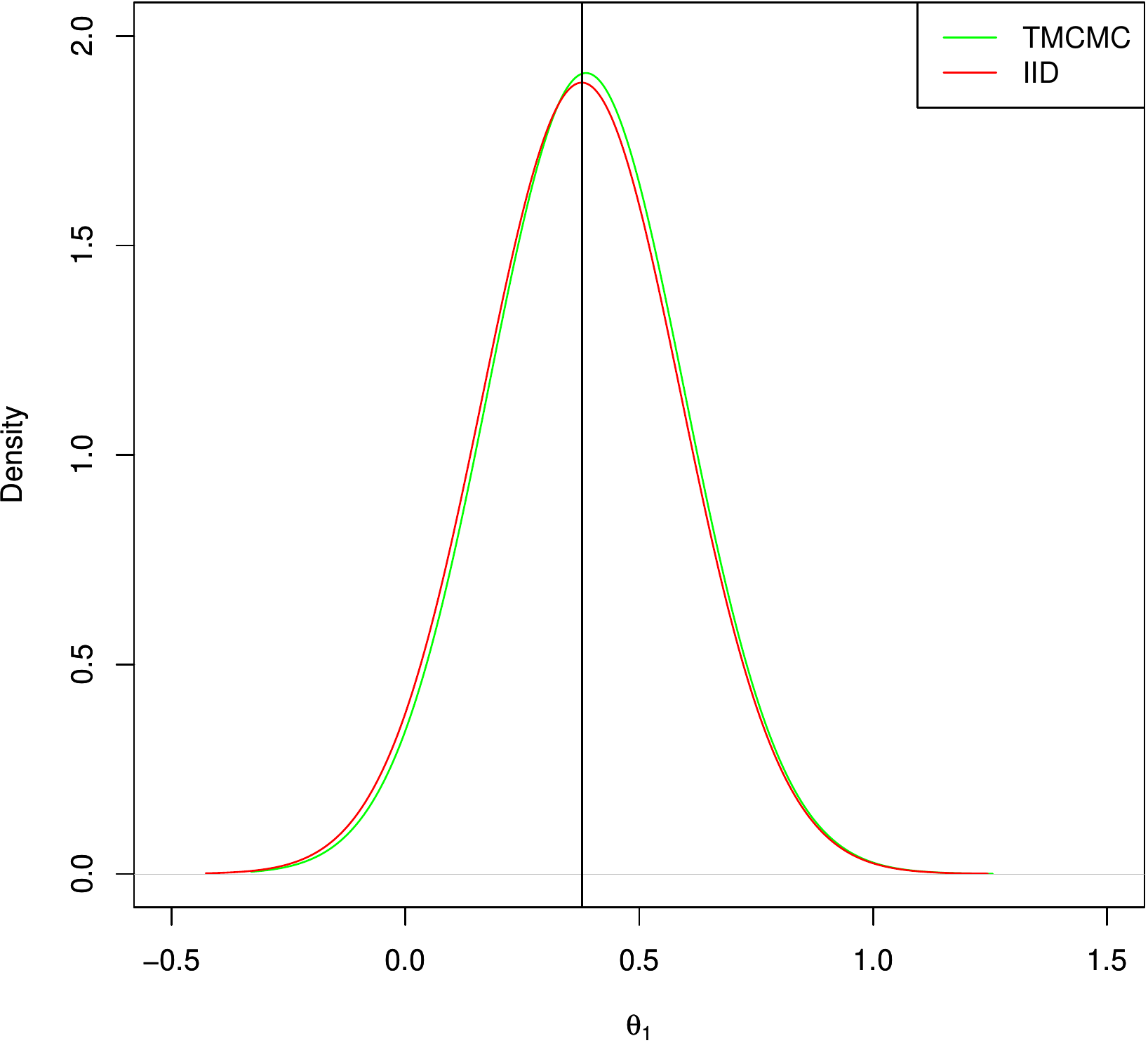}}
	\caption{Ising model simulation study. The green and red colours denote TMCMC and $iid$ based density estimates, respectively. The vertical lines
	stand for the true values.}
	\label{fig:ising}
\end{figure}

Since the true marginal posterior distributions are not available here (and generally for doubly intractable posteriors), 
Figure \ref{fig:ising} compares the marginal posterior distributions of $\theta_0$ and $\theta_1$ based on TMCMC and the $iid$ method.
The true, data-generating values of $\theta_0$ and $\theta_1$ are also presented as the vertical lines.
That the TMCMC solutions are very close to the $iid$ solutions show that even TMCMC is extremely reliable. That the true values of $\theta_0$ and $\theta_1$ fall in the
respective highest posterior density regions, also vindicate that the posteriors are reliably computed by our methods.

\subsection{Strauss process posterior}
\label{subsec:strauss}

The Strauss point process (\ctn{Strauss75}) on a bounded region $S\subset\mathbb R^2$ has unnormalized density
\begin{equation}
	f(\bY|\beta,\gamma)=\beta^{n}\gamma^{s_R(\bY)},
	\label{eq:strauss}
\end{equation}
with respect to the measure $\nu$, corresponding to the homogeneous Poisson point process on $S$ with unit intensity.
In (\ref{eq:strauss}), $\beta>0$, $0<\gamma\leq 1$, $\bY=\{\by_1,\ldots,\by_n\}$ is a point configuration where $\by_i\in S$ for $i=1,\ldots,n$, and 
$s_R(\by)=\sum_{i=1}^n\sum_{j>i}I_{\left\{\|\by_i-\by_j\|\leq R\right\}}(\by_i,\by_j)$. 

We assume {\it a priori}, as in \ctn{Moller06}, that $\beta\sim U(0,150)$ and $\gamma\sim U(0,1)$.

\subsubsection{Data generation}

With true values of $\beta$ and $\gamma$ set to be $100$ and $0.5$, respectively, we generated a realization of the Strauss process on $S=[0,1]^2$
with $R=0.05$ using the ``spatstat" package (\ctn{Baddeley05}) and obtained $n=78$ points.

\subsubsection{Importance sampling for obtaining $\bZ$}
For importance sampling, here we consider the density of the homogeneous Poisson point process with intensity $\rho=n$ as the envelope density on $S=[0,1]^2$.
The density is given by $g(\bX)=\exp\left\{(1-\rho)\right\}\rho^N$, with respect to $\nu$,
the homogeneous Poisson point process on $S$ with unit intensity. Simulation from this density is straightforward; we first simulate $N$ from the Poisson
distribution with mean $\rho$, and then given $N$, simulate $\bx_i$; $i=1,\ldots,N$, independently and uniformly from $S$. Given $\beta$ and $\gamma$, 
to compute the ratio $f/g$ for importance sampling, $\bY$ and $n$ in (\ref{eq:strauss}) must be replaced with $\bX$ and $N$, respectively, drawn from the envelope density $g$.
In our application, we set $\rho=n$ to ensure that the average number of points are the same for both $f$ and $g$.

We generate $10,000$ realizations from $g(\cdot)$ to estimate the unknown normalizing constant given each value of $(\beta,\gamma)$.
With the reparameterization
$\beta=150\exp(\theta_1)/\left(1+\exp(\theta_1)\right)$ and $\gamma=\exp(\theta_2)/\left(1+\exp(\theta_2)\right)$,
respectively, where $-\infty<\theta_1,\theta_2<\infty$, so that $\btheta=(\theta_1,\theta_2)$, as before,
we randomly select $\tilde\btheta_1,\ldots,\tilde\btheta_K$ uniformly from the ball of unit radius, with $K=100$. 

\subsubsection{TMCMC implementation to estimate $\bmu$ and $\bSigma$}
The TMCMC details for $\btheta$ remains the same as before, and once again exhibited excellent mixing properties. Using the $10,000$ TMCMC realizations
we estimated $\bmu$ and $\bSigma$ corresponding to $\btheta$.

\subsubsection{Implementation of the $iid$ method}
Here we set $\sqrt{c_1}=3$ and for $i=2,\ldots,M~(=15)$, $\sqrt{c_i}=\sqrt{c_1}+0.05\times (i-1)$. 
As before we considered $\tilde N_i=5000$ to be the Monte Carlo sample size drawn uniformly from $\bA_i$ to construct the estimators
$\widehat{\tilde\pi(\bA_i|\by)}$. 

With the above details, implementation of Algorithm \ref{algo:perfect} on $80$ cores of our VMWare took $1$ hour $1$ minute for this problem
to generate $10,000$ $iid$ realizations from the doubly intractable posterior.

\subsubsection{Results}

\begin{figure}
	\centering
	\subfigure [TMCMC and $iid$-based posterior of $\beta$.]{ \label{fig:beta}
	\includegraphics[width=7.5cm,height=7.5cm]{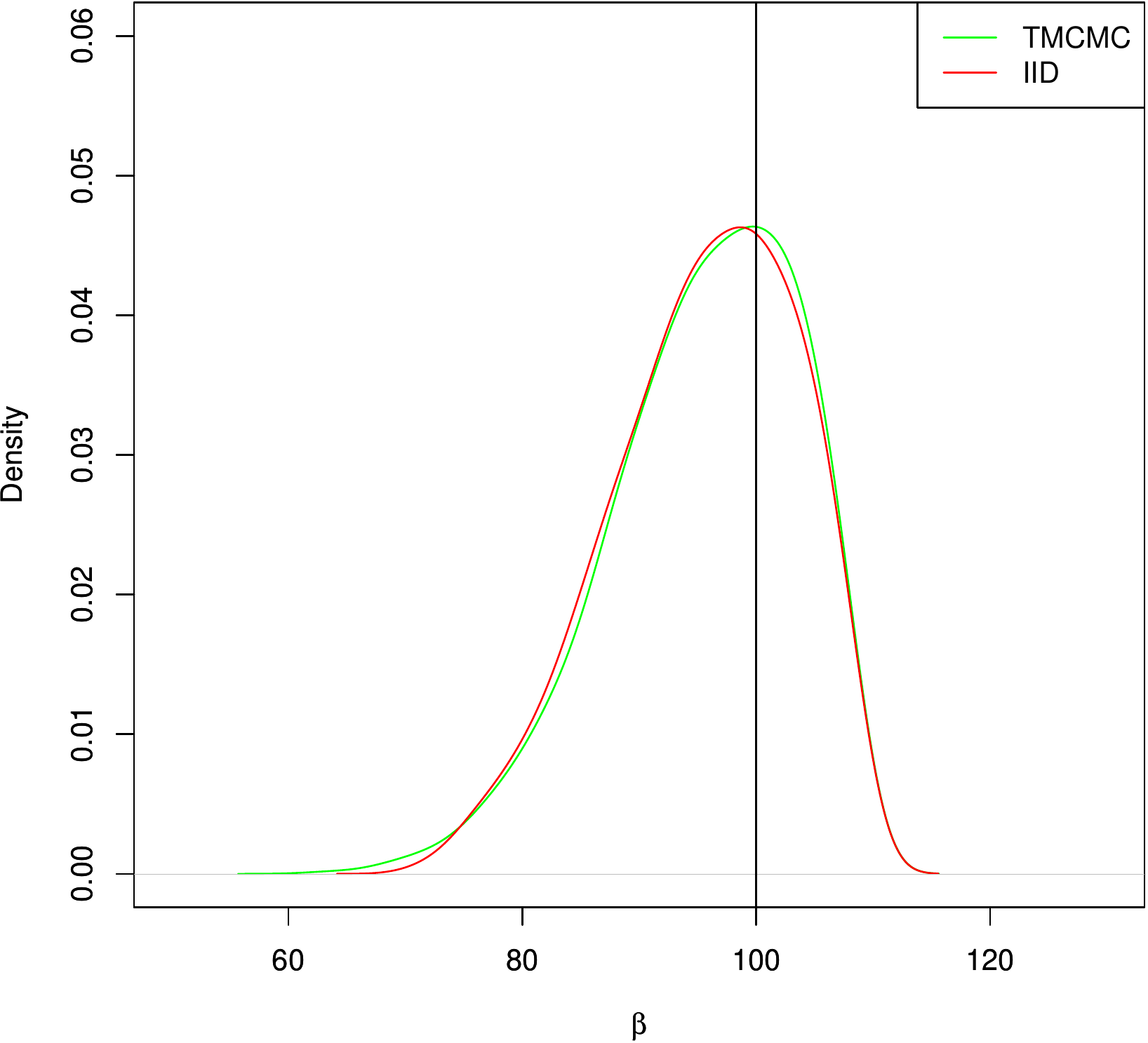}}
	\hspace{2mm}
	\subfigure [TMCMC and $iid$-based posterior of $\gamma$.]{ \label{fig:gamma}
	\includegraphics[width=7.5cm,height=7.5cm]{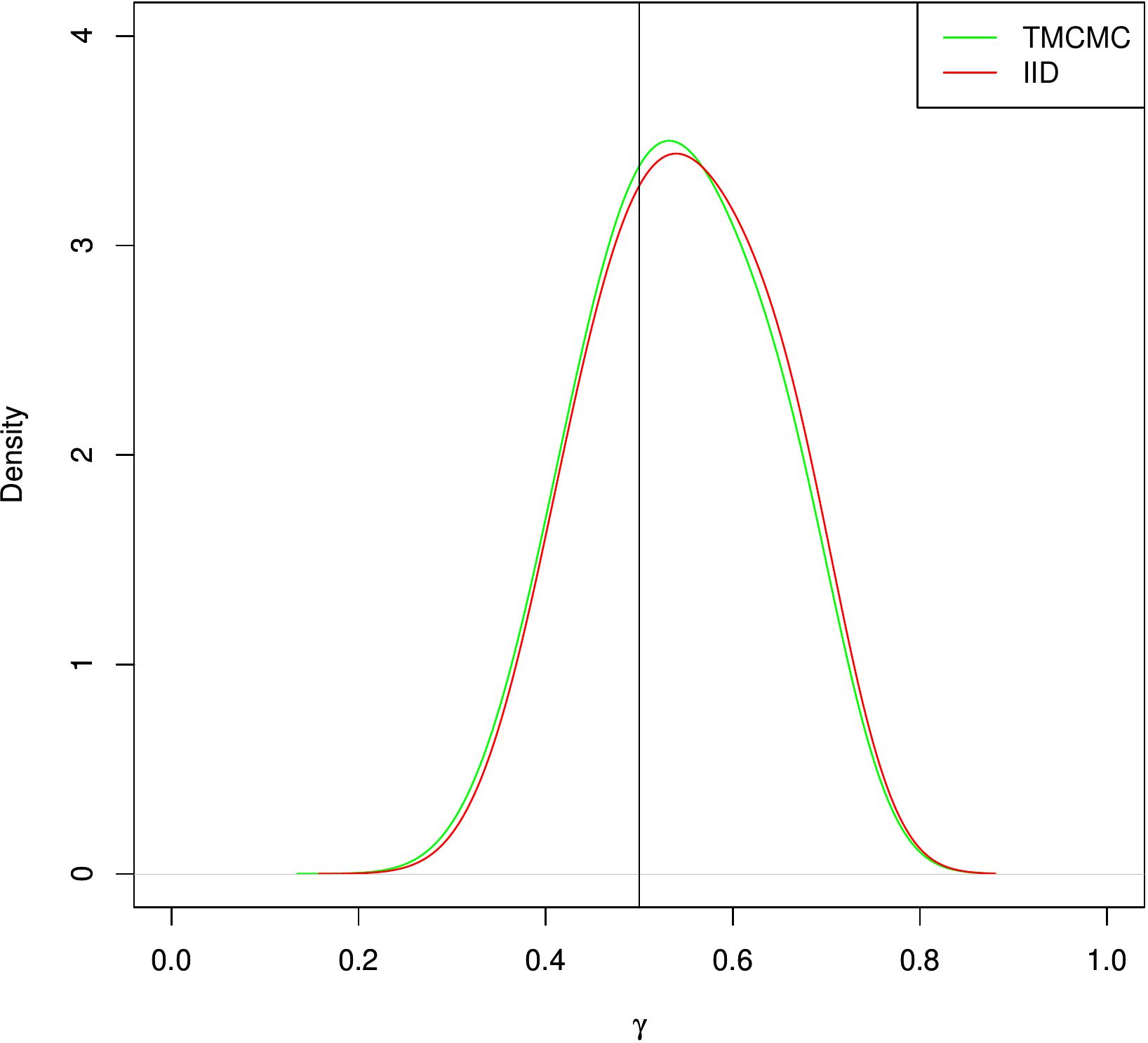}}
	\caption{Strauss process simulation study. The green and red colours denote TMCMC and $iid$ based density estimates, respectively. The vertical lines
	stand for the true values.}
	\label{fig:strauss}
\end{figure}

Figure \ref{fig:strauss} compares the marginal posterior distributions of $\beta$ and $\gamma$ based on TMCMC and the $iid$ method, where
the vertical lines represent the respective true, data-generating values.
Once again, very close agreement between the two methods, with inclusion of the true values in the highest density regions, vindicate their reliability and accuracy.

\subsection{Autologistic posterior}
An autologistic model for $\by=(y_1,\ldots,y_n)$, where $y_i\in\{-1,1\}$, has unnormalized density
\begin{equation}
	f(\by|\bphi)=\exp\left(\sum_{i=1}^n\phi_iy_i+\sum_{1\leq i<j\leq n}\phi_{i,j}y_iy_j\right),
	\label{eq:auto}
\end{equation}
where $\bphi$ consists of the real parameters $\phi_i$ and $\phi_{i,j}$ ($=\phi_{j,i}$). Note that (\ref{eq:auto}) generalizes the Ising model (\ref{eq:ising}).
See, for example, \ctn{Moller06} for further details. 

For our purpose, we set $\phi_i=\phi$ for $i=1,\ldots,n$ and $\phi_{i,j}=0$, whenever $j-i>1$. The model (\ref{eq:auto}) then reduces to the form
\begin{equation*}
	f(\by|\bxi)=\exp\left(\xi_1\sum_{i=1}^ny_i+\sum_{i=2}^n\xi_iy_{i-1}y_i\right),
\end{equation*}
where $\bxi=(\xi_1,\ldots,\xi_n)$.

For the prior on $\bxi$, we assume that $\xi_1\sim U(-1,1)$ and for $i\geq 2$, $\xi_i\sim U(0,1)$, independently.

\subsubsection{Data generation}
With $n=100$, we first generated the values of $\bxi$-components from their respective prior distributions. Then, given $\bxi$, we implemented
Gibbs sampling, run to $10^6$ iterations to store the $\by$ configuration of the last iteration as the generated data. In this regard, the full conditional distribution
of $y_i$ given $\by_{-i}=(y_1,\ldots,y_{i-1},y_{i+1},\ldots,y_n)$, is given by
\begin{equation}
	\left[\frac{\left(y_i+1\right)}{2}\bigg |\by_{-i}\right]\sim\mathcal B\left(p_i\right),
	\label{eq:auto_fullcond}
\end{equation}
where, 
\begin{equation*}
	p_i=\frac{\exp\left(\xi_1+\xi_{i+1}y_{i+1}I_{\left\{(i+1)\leq n\right\}}(i)+\xi_{i}y_{i-1}I_{\left\{(i-1)\geq 1\right\}}(i)\right)}
	{1+\exp\left(\xi_1+\xi_{i+1}y_{i+1}I_{\left\{(i+1)\leq n\right\}}(i)+\xi_{i}y_{i-1}I_{\left\{(i-1)\geq 1\right\}}(i)\right)}.
\end{equation*}

\subsubsection{Importance sampling for obtaining $\bZ$}
As before, we reparameterize the components of $\bxi$ such that $\xi_1=-1+2\exp(\theta_1)/(1+\exp(\theta_1))$ and $\xi_i=\exp(\theta_i)/(1+\exp(\theta_i))$, for $i\geq 2$,
where $-\infty<\theta_i<\infty$, for all $i$.
As in the case of Ising model, here we use the product of the full conditionals (\ref{eq:auto_fullcond}) given data $\by$ to construct the envelope distribution.
In other words, the envelope distribution in this case, given $\bxi$ (or $\btheta$) is
\begin{equation*}
	g(\bx|\bxi)=\prod_{i=1}^n\left[\frac{\left(x_i+1\right)}{2}\bigg |\by_{-i}\right],
\end{equation*}
where $\bx=(x_1,\ldots,x_n)$ and $\left[\frac{\left(x_i+1\right)}{2}\bigg |\by_{-i}\right]$ is of the same form as (\ref{eq:auto_fullcond}).
With this importance sampling distribution, we obtain $\bZ$ of size $K=500$ corresponding to $\tilde\btheta_1,\ldots,\tilde\btheta_K$ drawn uniformly and independently
from an $n$-dimensional ball with radius $2$. For the remaining $\btheta$ values we use (\ref{eq:normconst_gp}) for approximating the unknown normalizing constants.

\subsubsection{TMCMC implementation}
Here we implement a TMCMC algorithm that is a mixture of additive and multiplicative transformations (with probability $1/2$) 
with respect to $\btheta$, for this $n~(=100)$-dimensional problem. Further deterministic additive and multiplicative steps (with probability $1/2$)
are incorporated at the end of each iteration. For details regarding this algorithm, see Algorithm 2 of \ctn{Roy20}.
The other details remain the same as before. Again, our diagnostics revealed excellent performance of our algorithm, and the time taken remained less than a minute.
With the $10,000$ TMCMC realizations we estimated $\bmu$ and $\bSigma$ corresponding to the $100$-dimensional parameter $\btheta$.

\subsubsection{Implementation of the $iid$ method}
Here we set $\sqrt{c_1}=5$ and for $i=2,\ldots,M~(=10000)$, $\sqrt{c_i}=\sqrt{c_1}+0.0005\times (i-1)$. 
As before we considered $\tilde N_i=5000$ to be the Monte Carlo sample size drawn uniformly from $\bA_i$ to construct the estimators
$\widehat{\tilde\pi(\bA_i|\by)}$. 

However, this high-dimensional scenario required diffeomorphism to ensure significant minorization probabilities for perfect sampling. With $b=0.01$
as the diffeomorphism parameter, we implemented the idea outlined in Section \ref{subsec:diffeo} (see \ctn{Bhatta21a} for details) for $iid$ sampling
in conjunction with Algorithm \ref{algo:perfect}. We also evaluated $\widehat{\tilde\pi(\bA_i|\by)}$ using the diffeomorphism idea presented
in Section 3 of \ctn{Bhatta21b}, although the numerical values turned out to be extremely close to the non-diffeomorphed evaluations.

With the above details, implementation of our diffeomorphed $iid$ method on $80$ cores of our VMWare took just $2$ minutes for this high-dimensional problem
to generate $10,000$ $iid$ realizations.

\subsubsection{Results}

\begin{figure}
	\centering
	\subfigure [TMCMC and $iid$-based posterior of $\xi_1$.]{ \label{fig:xi1}
	\includegraphics[width=7.5cm,height=7.5cm]{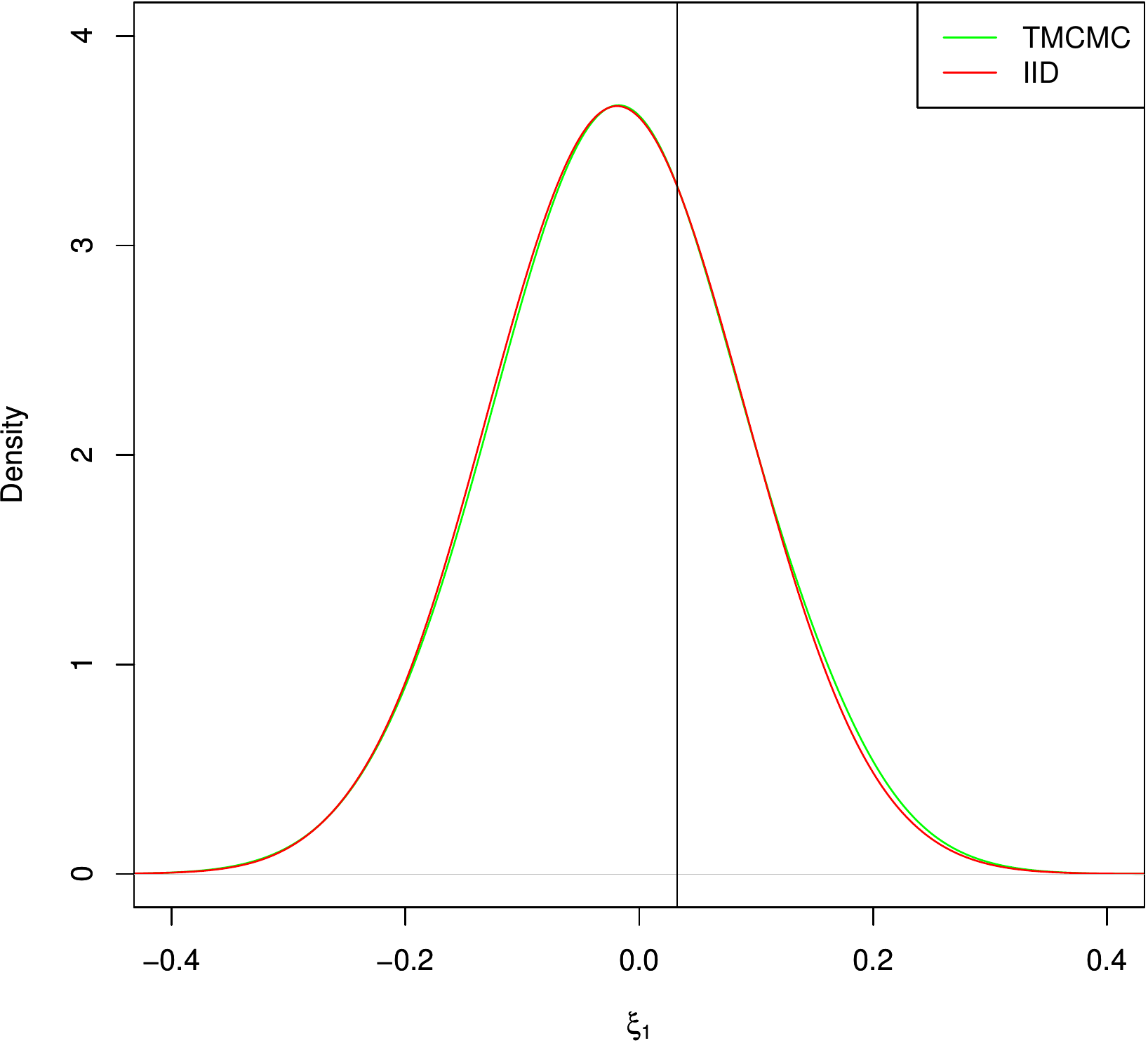}}
	\hspace{2mm}
	\subfigure [TMCMC and $iid$-based posterior of $\xi_{25}$.]{ \label{fig:xi25}
	\includegraphics[width=7.5cm,height=7.5cm]{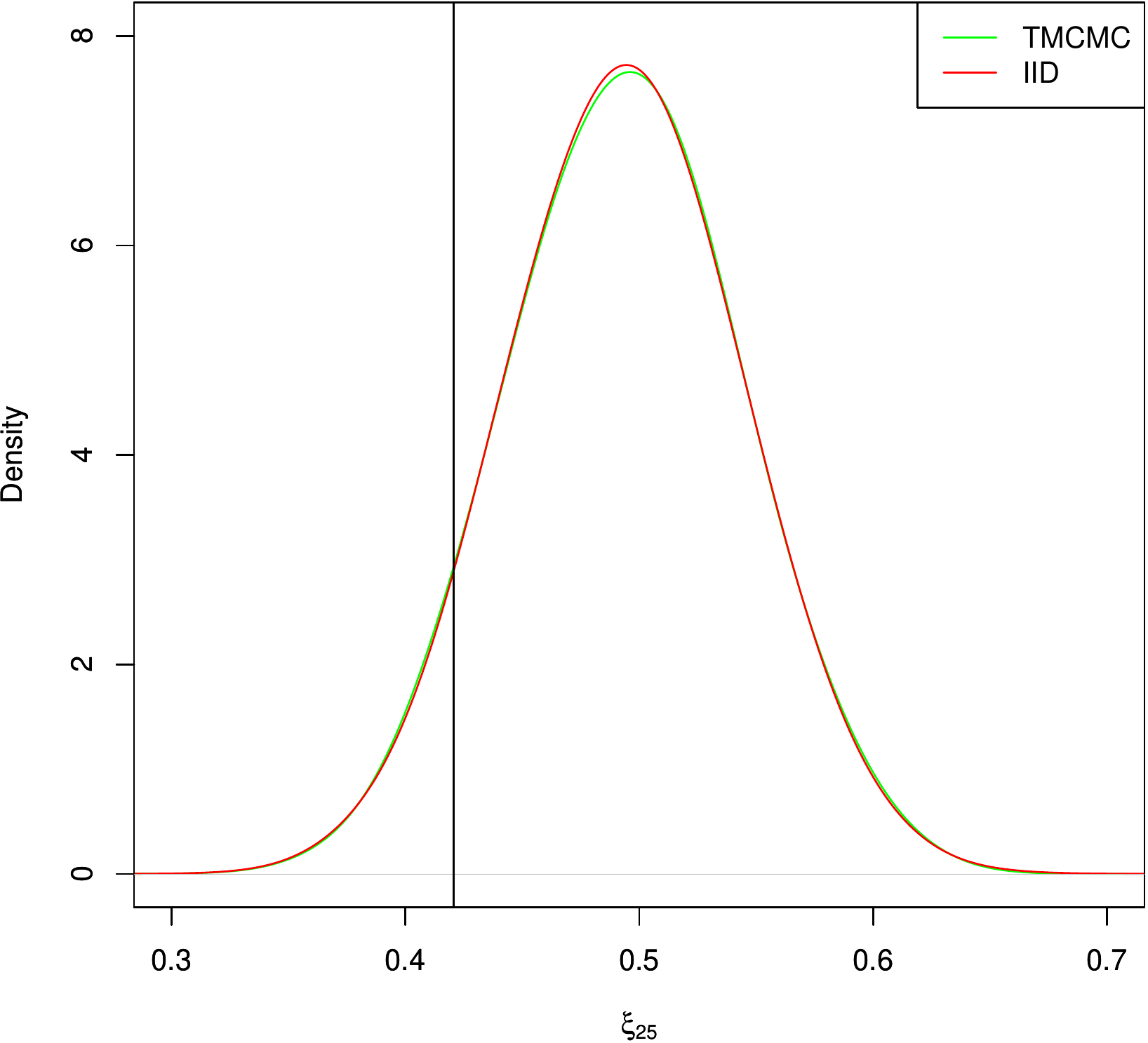}}\\
	\vspace{2mm}
	\subfigure [TMCMC and $iid$-based posterior of $\xi_{75}$.]{ \label{fig:xi75}
	\includegraphics[width=7.5cm,height=7.5cm]{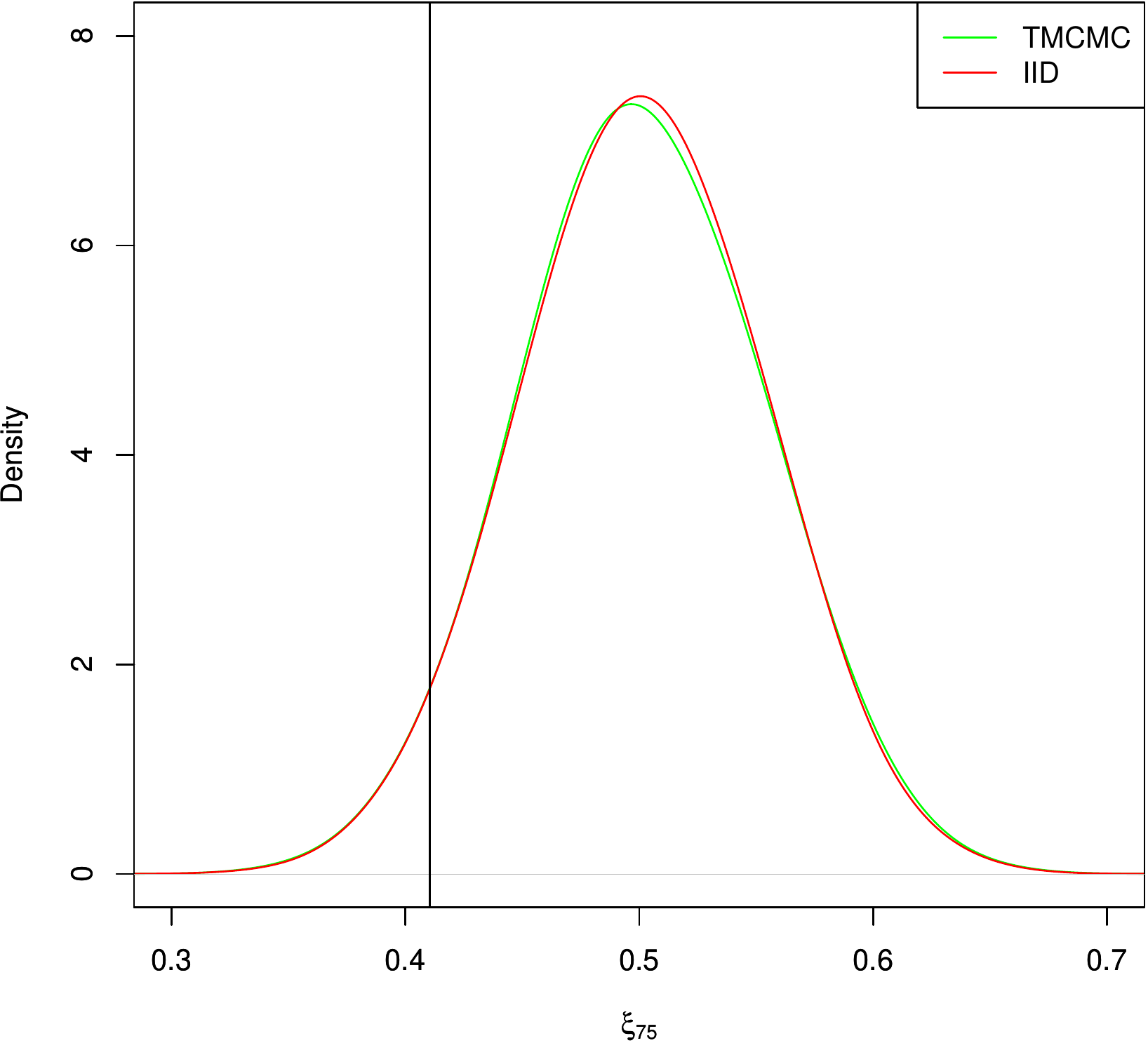}}
	\hspace{2mm}
	\subfigure [TMCMC and $iid$-based posterior of $\xi_{100}$.]{ \label{fig:xi100}
	\includegraphics[width=7.5cm,height=7.5cm]{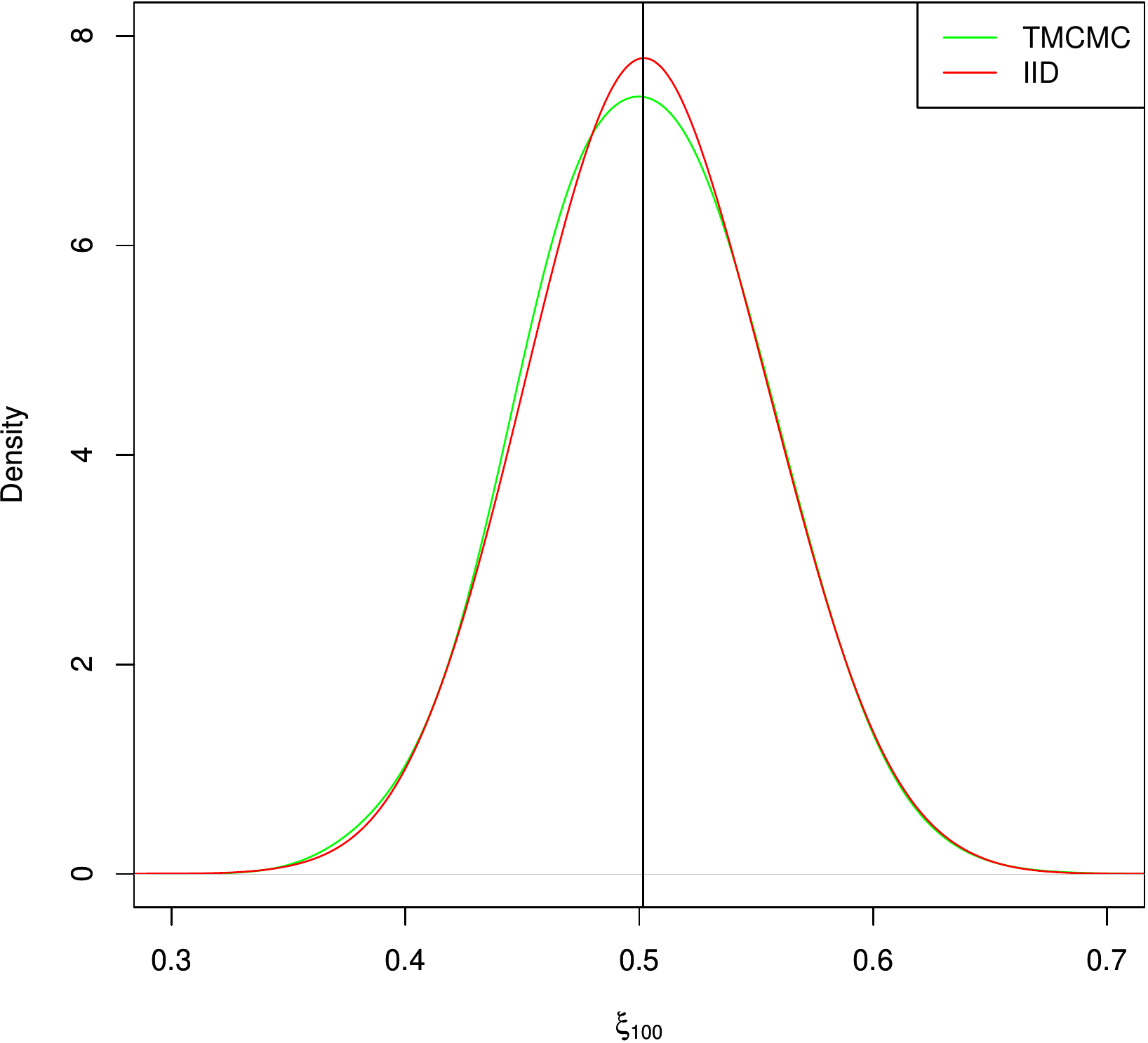}}\\
	\caption{Autologistic model simulation study. The green and red colours denote TMCMC and $iid$ based density estimates, respectively. The vertical lines
	stand for the true values.}
	\label{fig:auto}
\end{figure}

Figure \ref{fig:auto} compares the marginal posterior distributions of the components of $\bxi$ based on TMCMC and the $iid$ method, where
the vertical lines as before represent the respective true, data-generating values.
Again, as always, very close agreement between the two methods, with inclusion of the true values in the high density regions, vindicate their reliability and accuracy.

\section{Summary and conclusion}
\label{sec:conclusion}

Doubly intractable distributions constitute a very challenging area of Bayesian inference, where, in spite of the MCMC revolution, 
effective solutions are hard to obtain. The main issue is of course the parameter-dependent unknown normalizing constants that do not cancel in the
acceptance ratio of the Metropolis-Hastings algorithm. Attempts to provide valid MCMC algorithms for doubly intractable distributions mostly
rely upon specially constructed auxiliary variables to bypass evaluation of the normalizing constant, and are typically computationally burdensome
with no guarantee of good mixing properties. Usually, strong assumptions are also necessary for theoretical justification. 
Other methods attempt to approximate the normalizing constants by various means and need not converge to the exact target distribution.
The large body of ABC methods are very popular, but usually ad-hoc and unreliable. It is somewhat disconcerting to note that so far all the illustrations
of the existing methods for doubly intractable distributions considered just a few parameters (to our knowledge, not exceeding $10$).

In this article, following the $iid$ sampling development of \ctn{Bhatta21a} and \ctn{Bhatta21b}, we have attempted to create a novel methodology
for $iid$ sampling from doubly intractable distributions. The key is to approximate the unknown normalizing constant by an effective combination
of Monte Carlo/importance sampling and Gaussian process interpolation. The approximation can be made arbitrarily accurate in a manner that the 
realizations generated under the exact and approximated normalizing constants are the same, given the same algorithm and uniform random numbers are used for their generations.

Our illustrations with posteriors associated with a normal-gamma model, Ising model, Strauss process and a $100$-parameter autologistic model,
vindicated great accuracy and computational efficiency of our method. At least for the normal-gamma model, where the parameter-dependent normalizing 
constant is known but pretended to be unknown for validation of our method, the posterior is completely known analytically. For the other examples,
the posteriors are {\it bona fide} doubly intractable distributions. Thus, the normal-gamma example constitutes the benchmark for justification of our
$iid$ method. Very close agreement of our $iid$ based posteriors with the exact posterior distributions corresponding
to the normal-gamma example, is indeed a vindication of the triumph of our $iid$ methodology. For the rest of the examples, we compared our $iid$ method with
our fast and highly reliable TMCMC algorithms. Very close agreement between the two, along with inclusion of the true, data-generating
parameter values in the high-density regions, in all the relevant cases, even for the $100$-dimensional autologistic setup, where dffeomorphism based
implementation of our $iid$ algorithm was warranted, is very highly encouraging.
Very interestingly and encouragingly, all our Gaussian process interpolations required for approximating the normalizing constants, are based on
a small number of parameter values simulated from balls with appropriately chosen radii. Indeed, the normal-gamma, Ising and the Strauss examples
required only $100$ such values, whereas the $100$-dimensional autologistic example required only $500$. Moreover, these are embarrassingly parallelizable. 
Also recall that our Gaussian process interpolation formula is amenable to extremely efficient computation, since most of the computations can be done only once,
even before beginning implementation of the main $iid$ algorithm.
Needless to mention, these contributed to very
fast implementation of our algorithm, without compromising on accuracy. 

In this work, we have confined ourselves to applications based on simulated data. There are two reasons behind the choice of exclusively simulation experiments.
The first is the requirement of validation of our $iid$ method for which the true, data-generating values of the parameters are necessary, since agreement
of the $iid$ method with TMCMC need not be sufficient, as the latter is also based upon approximation of the normalizing constants. Inclusion of the true parameter values
in the high-density regions of the simulation based posteriors provides some testimony that the correct posteriors have been well-explored.
Secondly, we are not aware of any real data based doubly intractable posterior distribution of high-dimensional parameter vector, on which we can apply our method.
The autologistic model allowed us to simulate a high-dimensional scenario to illustrate the power of our creation.

In future, we shall apply our $iid$ idea to real life data associated with doubly intractable posteriors, with particular interest in high-dimensional scenarios.


\renewcommand\baselinestretch{1.3}
\normalsize
\bibliography{irmcmc}


\end{document}